\documentclass{article}
\usepackage{blindtext}
\usepackage[a4paper, total={6in, 8in}]{geometry}

\usepackage{cite}
\usepackage{amsmath,amssymb,amsfonts}
\usepackage{algorithmic}
\usepackage{graphicx}
\usepackage{textcomp}
\usepackage{xcolor}
\usepackage[utf8]{inputenc}
\usepackage[english]{babel}
\usepackage[T1]{fontenc}
\usepackage{amsmath}
\usepackage{hyperref}
\usepackage{float}
\usepackage{caption}
\usepackage{tikz}
\usepackage{lipsum}
\usepackage{comment}
\usepackage{float}

\usepackage[acronym]{glossaries}
\newacronym{gw}{GW}{Goemans-Williamson}
\newacronym{qaoa}{QAOA}{Quantum approximate optimization algorithm}
\newacronym{qaoa2}{QAOA$^2$}{QAOA-in-QAOA}
\newacronym{mpi}{MPI}{Message Passing Interface}
\newacronym{vqa}{VQA}{Variational Quantum Algorithm}
\newacronym{nisq}{NISQ}{Noisy Intermediate-Scale Quantum}
\newacronym{maxcut}{MaxCut}{Maximum Cut}
\newacronym{qc}{QC}{Quantum computation}
\newacronym{qubo}{QUBO}{quadratic unconstraing binary optimization}
\newacronym{mpmd}{MPMD}{Multiple-Program-Multiple-Data}
\newacronym{sdp}{SDP}{semi-definite rogramming}
\newacronym{ml}{ML}{machine learning}
\newacronym{hpc}{HPC}{high performance computing}
\newacronym{scs}{SCS}{splitting conic solver}

\usepackage{physics}
\usepackage{graphicx}
\usepackage{caption}
\usepackage{subcaption}
\graphicspath{ {./images/} }
\usepackage{xcolor}
\usepackage[normalem]{ulem}

\usepackage{xcolor}     
\usepackage[normalem]{ulem}
\usepackage{enumitem}
\usepackage{authblk}

\usepackage{mathtools}
\DeclarePairedDelimiter\ceil{\lceil}{\rceil}

\newcommand{\betagamma}{(\vec{\beta},\vec{\gamma})} 
\def\BibTeX{{\rm B\kern-.05em{\sc i\kern-.025em b}\kern-.08em
    T\kern-.1667em\lower.7ex\hbox{E}\kern-.125emX}}

\newcommand\blfootnote[1]{%
  \begingroup
  \renewcommand\thefootnote{}\footnote{#1}%
  \addtocounter{footnote}{-1}%
  \endgroup
}

\providecommand{\keywords}[1]
{
  \small	
  \textbf{\textit{Keywords---}} #1
}

\title{Hybrid Classical-Quantum Simulation of MaxCut using QAOA-in-QAOA}
\author[1]{Aniello Esposito*}
\author[2]{Tamuz Danzig*}
\affil[1]{\textit{EMEA Research Lab}, \textit{Hewlett Packard Enterprise}, Basel (Switzerland), \texttt{aniello.esposito@hpe.com}}
\affil[2]{\textit{Classiq Technologies}, Tel Aviv (Israel), \texttt{tamuz@classiq.io}}
\begin{document}
\maketitle
\begin{abstract}
The \gls{qaoa} is a leading hybrid classical-quantum algorithm for solving complex combinatorial optimization problems. \gls{qaoa2} uses a divide-and-conquer heuristic to solve large-scale \gls{maxcut} problems, where many sub-graph problems can be solved in parallel.
In this work, an implementation of the \gls{qaoa2} method for the scalable solution
of the \gls{maxcut} problem is presented, based on the Classiq platform. 
The framework is executed on an HPE-Cray EX supercomputer by means of the \gls{mpi}
and the SLURM workload manager.
The limits of the \gls{gw} algorithm as a purely classical alternative to QAOA
are investigated to understand if \gls{qaoa2} could benefit from solving certain sub-graphs classically.
Results from large-scale simulations of up to 33 qubits are presented, showing the advantage of \gls{qaoa} in certain cases and the efficiency of the implementation, as well as the adequacy of the  workflow in the preparation of real quantum devices. 
For the considered graphs, the best choice for the sub-graphs does not significantly improve results and is still outperformed by \gls{gw}. \blfootnote{* Equal contribution.}
%
\end{abstract}
\newpage
\tableofcontents
\newpage
\section{Introduction}
\gls{qc} is an active area of research that has attracted key players in academia and industry, primarily due to its potential to solve specific problems much more efficiently that their classical counterparts \cite{nielsen2010quantum, arute2019quantum}. However, current \gls{nisq} devices feature a modest number of qubits and useful compute time is limited due to the decoherence \cite{sharma2020noise, stein2022eqc}. This has driven the development of methods to partition quantum circuits whose components are executed
on smaller quantum devices with assistance of classical processing \cite{de2023hybrid, huembeli2022entanglement, gentinetta2023overhead, brenner2023optimal, piveteau2023circuit, mitarai2021overhead, mitarai2021constructing, peng2020simulating, bravyi2016trading}. 
These methods can be roughly subdivided in circuit knitting and entanglement forging and are
naturally implemented as hybrid classical-quantum workflows. While quantum devices are expected
to perform well for certain problems involving a small amount of data but high complexity, 
current classical supercomputers continue to serve researchers in areas with data-intensive 
workloads relying on software tools and infrastructure developed over decades. Therefore, in
conjunction with a number of possibly tightly integrated \gls{nisq} devices, classical supercomputers 
would be valuable candidates for the execution of hybrid workflows while sustaining the classical portion of the computation and communication of potentially large data volumes. 
However, while considerable resources have been invested in quantum circuit simulators to accompany the 
development of quantum hardware, the investigation of \gls{hpc} systems featuring quantum hardware 
has though driven programming environments\cite{kim2023cuda, qaptiva, wu2023intel,lee2023quantum} but other operational aspects such as workflows\cite{esposito2023hybrid,  elsharkawy2023challenges, farooqi2023exploring, mete2022predicting, schulz2023towards, ruefenacht2022bringing, schulz2022accelerating, da2023workflows} have not received a comparable attention. 

This work focuses on the hybrid workflow, which is investigated by means of the SLURM\cite{yoo2003slurm}
workload manager on a HPE-Cray EX supercomputer which executes quantum circuits on simulated devices. 
The case study consists of 
a \gls{maxcut} problem, which has gained popularity in the \gls{qc} community 
due to its natural mapping to qubit measurement results. The solution can be approximated by the \gls{qaoa}\cite{farhi2014quantum} on quantum devices for general quantum circuit execution\cite{shaydulin2023qaoa} or conversely formulated as \gls{qubo} problem and solved with quantum annealers\cite{van2022evaluating}. The 
most popular classical alternative for approximate solutions is given by the \gls{gw} algorithm. 
Partitioning schemes for the \gls{maxcut} problem and \gls{qaoa} have been extensively investigated, either targeting the generated quantum circuit\cite{lowe2023fast, bechtold2023investigating} or 
the original classical graph\cite{zhou2023qaoa}, i.e. \gls{qaoa2}, using a divide-and-conquer heuristic. The latter approach is considered in this work and implemented with the support of the 
\texttt{Classiq platform} \cite{classiq}, being able to synthesize more optimized quantum 
circuits compared to a manual construction. Circuits are executed with 
the \texttt{aer} simulator \cite{qiskit-aer} using distributed memory parallelism over \gls{mpi}. This workflow naturally maps to the 
\gls{mpmd} or heterogeneous jobs paradigm of SLURM. While \gls{qaoa2} considers only \gls{qaoa}
for the sub-graphs, the SLURM features allows the allocation of a mixture of quantum and classical resources giving a choice between \gls{qaoa} or a classical solution of the sub-graph problem, for instance, with \gls{gw} to maximize the overall \gls{maxcut}. 
To that end, both methods are applied to
a set of graphs to assess the suitability of such a run-time decision mechanism. The present
implementation also provides a testbed to further refine such a mechanism and possibly include 
\gls{ml} methods similar to what has been considered in \cite{moussa2020quantum}.  

The paper starts with a consideration of related work, followed by details and remarks about the methods. Comparisons between the methods for moderate graph sizes are presented before the final method is applied to larger graphs. Finally, concluding remarks and an outlook are given.

\section{Related Work}
%
%
Hybrid classical-quantum workflows have been investigated\cite{esposito2023hybrid,  elsharkawy2023challenges, farooqi2023exploring, mete2022predicting, schulz2023towards, ruefenacht2022bringing, schulz2022accelerating, da2023workflows} considering cloud environments and
more tightly integrated infrastructures. A number of expectations for components such as workload 
managers or circuit depths have emerged with related challenges. In this work, the ingredients
of the workflow are kept simple, using a state-of-the-art workload manager and circuit simulator in 
order to focus on the suitability of a supercomputer by means of a well-known example application,
where the circuit depth is optimized by means of the \texttt{Classiq} platform. 
Partitioning of the \gls{maxcut} problem with circuit knitting of the resulting \gls{qaoa} formulation has also been investigated\cite{lowe2023fast, bechtold2023investigating}. However,
while this approach allows, in principle, the use of multiple quantum devices simultaneously, it 
discards the possibility to employ classical methods for a portion of the problem where it 
could be more advantageous. In this work, the \gls{qaoa2} method is considered as a more versatile 
example of hybrid application, where the \gls{gw} method \cite{goemans1995improved} is used for 
the classical solution of a sub-graph if more advantageous.

\gls{ml} methods have been considered by Moussa et al.~\cite{moussa2020quantum} who created a  classifier to predict the suitability of a graph instance for either\gls{qaoa} or \gls{gw}. They achieved 96\% prediction accuracy, though not at the qubit counts considered in this work. Their approach could be applied to a \gls{qaoa2} context as a next step. The present work provides a testbed to train and test such selection mechanisms. Furthermore, with a large dataset of \gls{qaoa} results, a neural network can be trained to predict initial parameters for subsequent \gls{qaoa} simulations or computations on real quantum hardware. This could improve the number of iterations in the hybrid scheme of \gls{qaoa} while preserving the accuracy \cite{amosy2022iterative}. 

Finally, it is worth mentioning that exact classical methods to solve the \gls{maxcut} have been 
extensively investigated\cite{rehfeldt2023faster} but the node counts in typical applications are 
still limited compared to \gls{gw}. Also probabilistic methods from statistical physics such as 
simulated annealing~\cite{kirkpatrick1983optimization} have been considered for \gls{maxcut}.
%
%
%
%
%
%
%
%
%
\section{Methods}
%
%
%
%
\subsection{The \gls{maxcut} problem}\label{Sec:MaxCut}
A graph $G=(V,E)$ is considered, with a set of nodes $V={1,2,...,N}$, and a set of edges $(i,j)\in E$ between the nodes $i$ and $j$ with weights $w_{ij} = w_{ji}$. The goal is to divide the nodes into two groups in such a way that maximizes the sum of weights of edges that touch nodes from the two different groups (cut) rather than the same group. The problem represents a NP-hard combinatorial problem.

The problem Hamiltonian for a given graph is:
%
\begin{equation}\label{eq:MaxCut_Cost}
H_C = \frac{1}{2}\sum_{(i,j)\in E}  w_{ij}\left(1-Z_{i}Z_{j}\right) \quad,
\end{equation}
where each qubit represents a node, and $Z_{i}$ is the Pauli operator that acts on qubit $i$. If the two qubits of the edge $(i,j)$ are measured differently, it adds to the overall sum. If they are measured similarly, there is no contribution to the sum.

\subsection{Quantum Approximate Optimization Algorithm}
\label{Sec:QAOA}
The method belongs to the family of \gls{vqa} and makes use of a specific parameterized quantum circuit structure (ansatz) \cite{farhi2014quantum} composed of $p$-layers
\begin{equation}\label{eq:QAOA_Ansatz}
\ket{\psi_{p}\betagamma} = \left[\prod_{l=1} ^p e^{-i\beta_{l} H_M} e^{-i\gamma_{l} H_C}\right] \ket{+}^{\otimes n},
\end{equation}
where $H_C$ is called the problem Hamiltonian, which is defined by the specific problem to be solved, $H_M$ is the mixer Hamiltonian that takes the same structure for any given problem, $n$ is the number of qubits in the quantum circuit, and $\vec{\gamma}$ and $\vec{\beta}$ are the variational real parameters that correspond to $H_C$ and $H_M$, respectively. The objective function is given by
\begin{equation}\label{eq:QAOA_Cost}
F_{p}\betagamma = \bra{\psi_{p}\betagamma} H_C \ket{\psi_{p}\betagamma} \quad, 
\end{equation}
which needs to be maximized (or minimized) by changing $\vec{\gamma}$ and $\vec{\beta}$ values in each iteration by a classical optimizer. The number of shots for the circuit simulation used in this
work is $4096$. 
The rigid structure of the Ansatz can be considered as a time-discretization of adiabatic quantum computation ~\cite{farhi2014quantum, crooks2018performance,farhi2000quantum}, and as $p \rightarrow \infty$, \gls{qaoa} obtain the optimal solution \cite{farhi2014quantum}.
The \gls{maxcut} problem has become the de-facto standard benchmark for \gls{qaoa} due to its simple formulation and suitability for binary encoding~\cite{zhou2020quantum, willsch2020benchmarking}.
%
Several works claim that \gls{qaoa} has the potential to achieve quantum advantage. Farhi and Harrow argue that \gls{qaoa} can reach quantum advantage based on complexity theoretic assumptions \cite{farhi2016quantum}. Guerreschi claims that \textit{"QAOA for MaxCut requires hundreds of qubits
for quantum speed-up"} \cite{guerreschi2019qaoa}. The number of qubits indeed grows in that direction. Crooks shows that when the number of layers is more than $8$, \gls{qaoa} achieves better results than \gls{gw} \cite{crooks2018performance}. Shaydulin et al. found that \gls{qaoa} gives the best empirical scaling of any known classical algorithm for the low auto-correlation binary sequence problem \cite{shaydulin2023evidence}. Bravyi et al. proposed a non-local variation of \gls{qaoa}, called recursive-QAOA (RQAOA), which numerically outperforms standard \gls{qaoa} \cite{bravyi2020obstacles}. This method can also be leveraged using \gls{qaoa2} to get a good global solution for very large problems in a reasonable time. The \gls{vqa} family of algorithms are especially suitable for the \gls{nisq} era because they use shallow quantum circuits with the assistance of classical computers \cite{preskill2018quantum, bharti2022noisy}. 
In this work, once the \gls{qaoa} circuit is executed, the bit string corresponding to the highest amplitude in the resulting state vector is chosen as a solution for sake of simplicity. A more appropriate approach would have been to consider a number of highest amplitudes and chose the bit string yielding the highest cut among them.
\subsection{QAOA-in-QAOA}
\label{Sec:QAOA2}
%
%
The method proposed by Zhou et al.~\cite{zhou2023qaoa} revisits the MaxCut problem via the divide-and-conquer heuristic by seeking the solutions of sub-graphs in parallel and then merging these 
solutions to obtain the global solution. It consists of the following steps.
\begin{enumerate}
\item Specify the number of qubits $n$ in our quantum computers or simulators, the number of layers in the quantum circuit, and the number of iterations. 
\item For the dividing procedure, the graph $G$ is partitioned into sub-graphs in which the number of nodes does not exceed a specified number of qubits $n$ from the previous step. The greedy modularity method from the \texttt{NetworkX} Library is used, which maximizes the modularity of the graph. If a sub-graph has more nodes than $n$, the sub-graph is divided into fewer sub-graphs, recursively.
\item All sub-graphs are solved with \gls{qaoa} in parallel over different (simulated) quantum devices. 
\item In the merging process, the overall solutions of all sub-graphs are considered. In order to obtain a better total solution, a new graph is generated as follows:
\begin{enumerate}[label=(\alph*)]
    \item Each sub-graph is represented by a node.
    \item Each edge that connects between two sub-graphs and is also part of the cut, we multiply its weight with $-1$. Each edge that is not part of the cut, we don't change.
    \item Take the sum on all edges between each two sub-graphs to add an edge with a single weight to the new graph. 
\end{enumerate}
\item The resulting graph is solved with \gls{qaoa}. If a node in the new graph is $-1$, all the nodes in the sub-graph represented by this node are flipped. If the new graph has more nodes than in the quantum computer, the process is repeated recursively.
%
%
%
%
%
\end{enumerate}
%
%
%
A total of $\sim \sum_{k=1}^{a} N / n^{k} = N(n^a-1)/(n^a(n-1))$ graphs have to be considered for $a$ levels. The condition $\ceil{N / n^a} \approx n$, i.e. that the lowest level graph is computed with $n$ qubits, yields $a\approx \ceil{log_{n} N} - 1$. If sufficient quantum devices are available to be used in parallel, the complexity reduces to $\sim O(\log_{n} N)$.
The original work shows experimentally that for different types of graphs, \gls{qaoa2} achieves competitive or even better results compared to \gls{gw} for 2000 nodes.
On the theoretical side, the algorithm is proven to be better than a random solution MaxCut, and a lower bound to the approximation ratio of \gls{qaoa2} is given, which depends on the partition method and the local solver. 
\subsection{Goemans-Williamson Algorithm}
It is the best-known classical approximation algorithm for the \gls{maxcut} problem, with an approximation ratio of $0.878$. 
The computationally most intensive part is the optimization of a \gls{sdp}
problem. Once the \gls{sdp} is solved, a slicing to determine the node values is applied $30$ times, and the average value of the cut is taken. This provides a more suitable way to compare to \gls{qaoa}, which is not repeated and averaged in the present work. For dense problems, the time to solution and memory consumption of \gls{gw} grow like $O(N^{6.5})$ and $O(N^{4})$\cite{zhang2018sparse}, respectively, where $N$ is the number of nodes in the 
graph.  The \texttt{cvxpy}
and \texttt{networkx} Python packages have been used for the solution of the \gls{sdp} problem and the graphs, respectively. The \texttt{cvxpy} disposes of a partial \texttt{OpenMP}
shared memory parallelization via the \gls{scs} in addition to the underlying threaded \texttt{BLAS/LAPACK} libraries but the scalability is rather limited and therefore only one \texttt{OpenMP} thread is used.
\subsection{Circuit Synthesis}
The \texttt{Classiq} platform \cite{classiq} allows a user to build a high-level functional quantum model. The synthesis engine takes this model along with optimization preferences and global constraints to produce an optimized quantum circuit while considering many different implementations. The synthesis engine can optimize over circuit depth, number of qubits, two-qubit gates, or create an optimized implementation into a specific hardware architecture, etc. In the present case, the description of a high-level combinatorial optimization problem is converted into an optimized gate-level quantum circuit.
\subsection{Workflow Execution}
\label{subsec:workflow}
\begin{figure}[htbp]
\centerline{\includegraphics[width=0.75\textwidth]{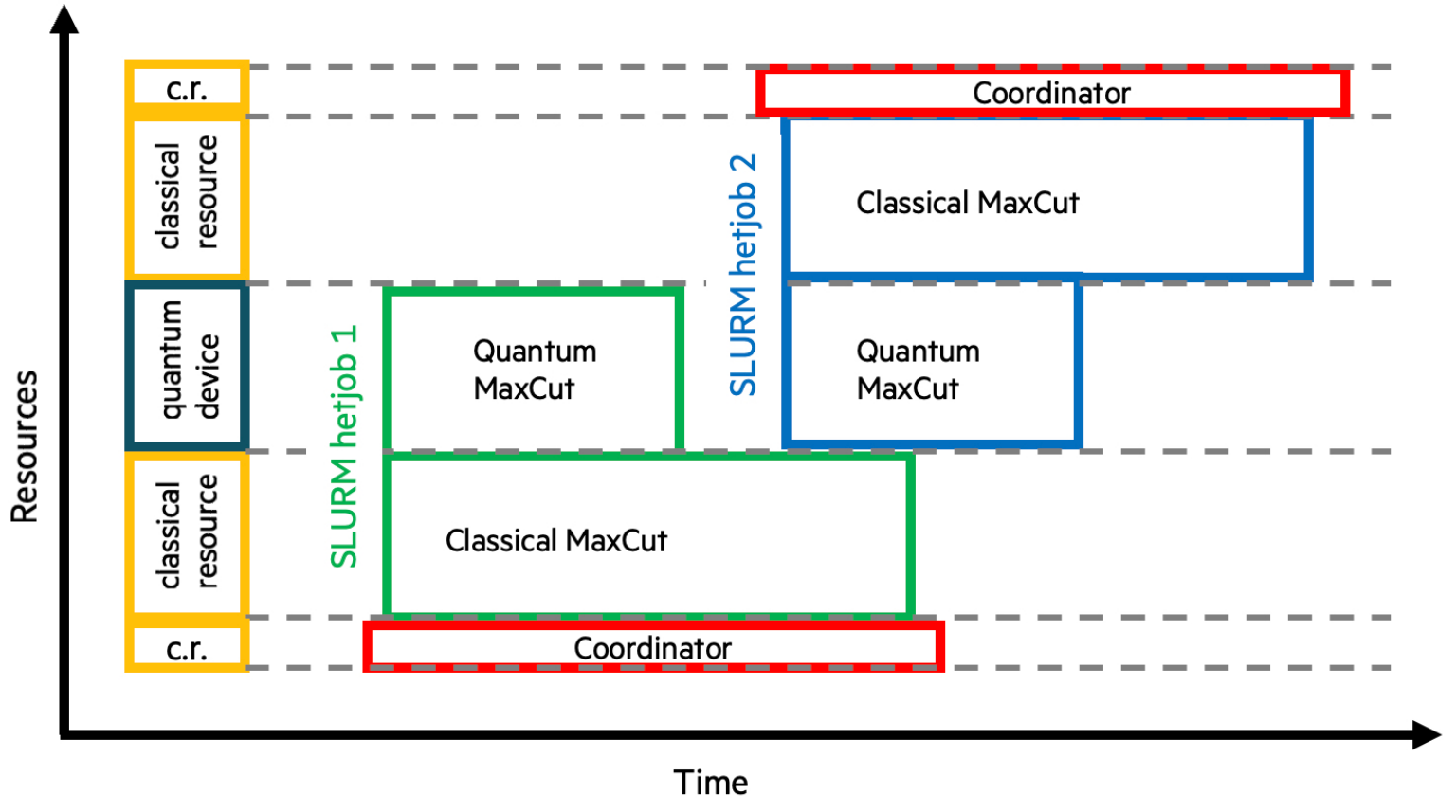}}
\caption{Heterogeneous jobs for the reduction of idle time of a quantum device. Before the first heterogeneous job finishes, a second one can already start using the quantum device. In this scenario, the quantum device is always accessed exclusively by a single user.}
\label{fig:het_jobs}
\end{figure}

\begin{figure*}[htbp]
\centerline{\includegraphics[width=0.975\linewidth]{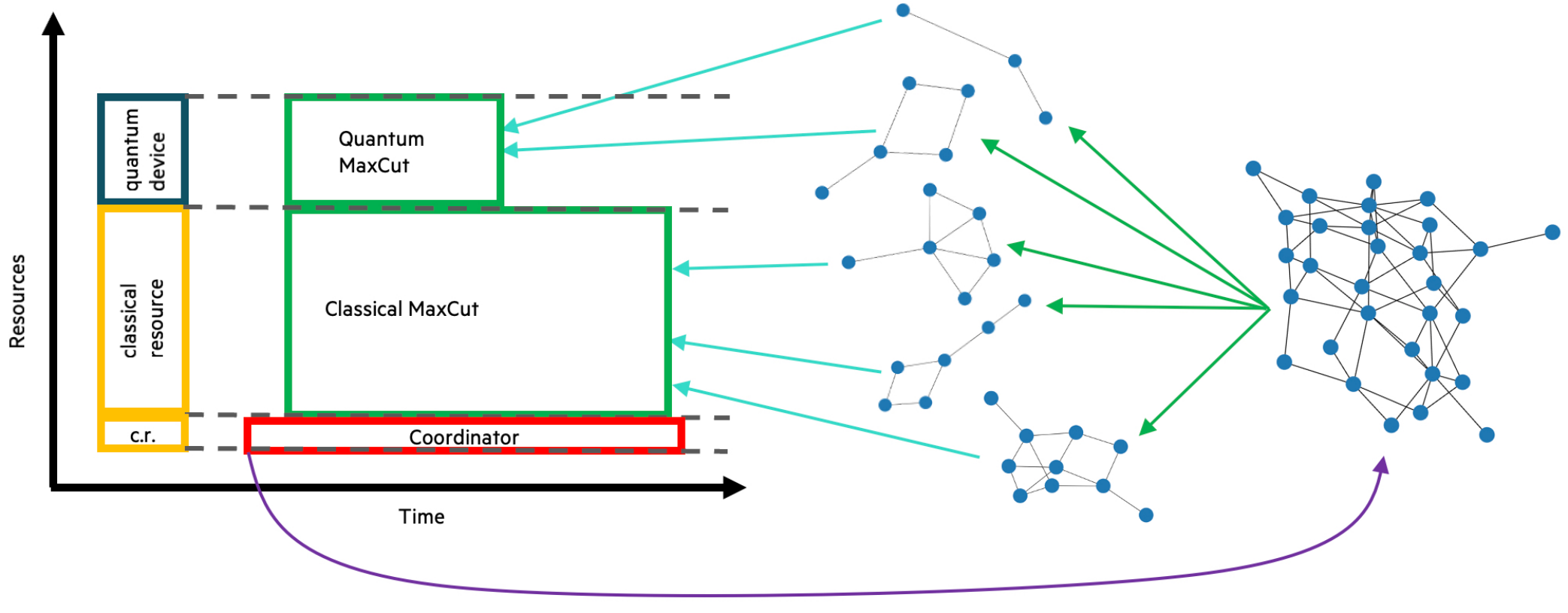}}
\caption{Distribution scheme for the solution of the sub-graphs generated by \gls{qaoa2}, either classically or quantum mechanically (simulated). A coordinator executed on a dedicated MPI rank handles the partitioning and collection of results. Usually, the consumption of classical and quantum resources does not start at the same time. However this can be achieved by splitting, checkpointing, and restarting the classical part appropriately.}
\label{fig:distribution_scheme}
\end{figure*}

The \gls{mpmd} paradigm in SLURM can be used to compute the solutions of the \gls{maxcut} problems of the sub-graphs generated by \gls{qaoa2}, either classically by the \gls{gw} method or quantum mechanically (simulated) by the \gls{qaoa} algorithm. This distinction is only made for
the first partitioning of the original graph in \gls{qaoa2} for the sake of simplicity in this preliminary investigation. 
A schematic view of the procedure is given in Fig.~\ref{fig:distribution_scheme}, where 
a coordinator could inspect the sub-graphs and calculate the most appropriate resource allocation in advance.
Furthermore, in order to save idle time caused by a possibly scarce usage of a quantum device, the heterogeneous jobs paradigm of SLURM could be employed, as shown in Fig.~\ref{fig:het_jobs}.
%
%
%
An HPE-Cray EX supercomputer featuring compute nodes with two AMD EPYC 7763 64-Core processors and 512 GB of memory is used for the experiments. Distributed memory parallelization is achieved through the \texttt{mpi4py}\cite{dalcin2021mpi4py} package. 
\section{Results}
\begin{figure*}[htbp]
    \centering
    \begin{subfigure}[b]{\textwidth}
        \centering
        \includegraphics[width=0.475\linewidth]{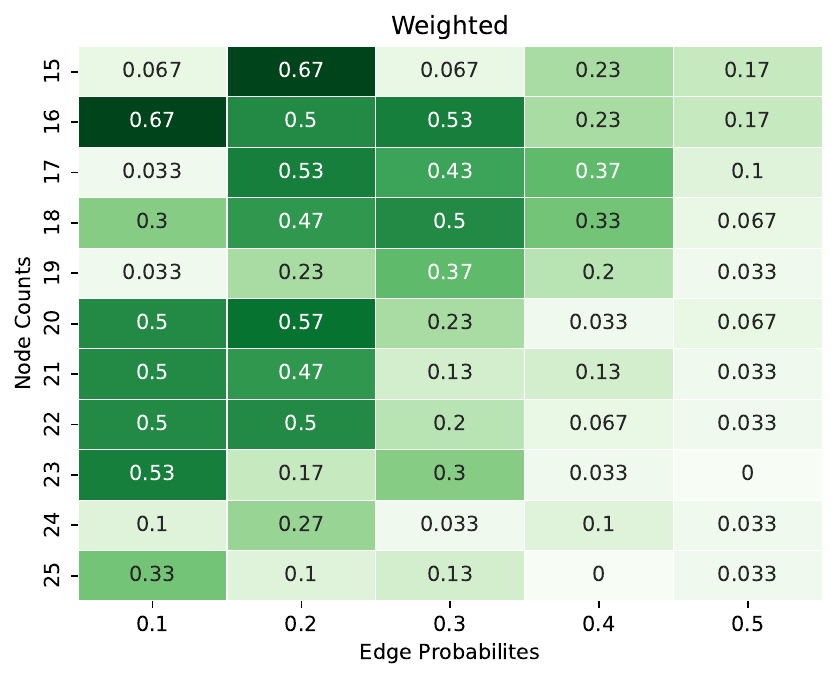}
    \hfill
        \includegraphics[width=0.475\linewidth]{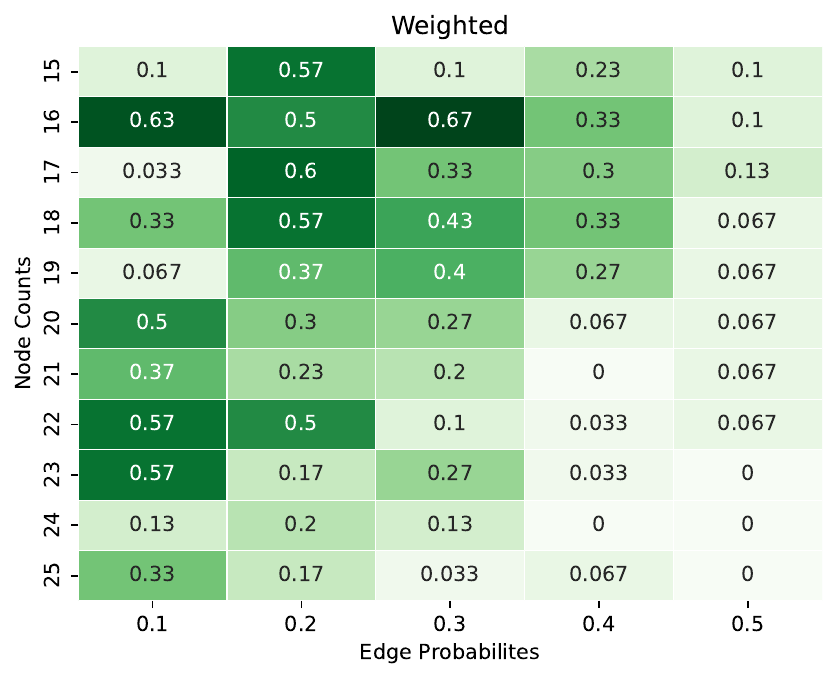}
        \caption{Proportions of cases in which \gls{qaoa} is strictly better than \gls{gw} for unweighted (left) and weighted (right) graphs.}   
    \end{subfigure}
    \vskip\baselineskip
    \begin{subfigure}[b]{\textwidth}   
        \centering 
        \includegraphics[width=0.475\linewidth]{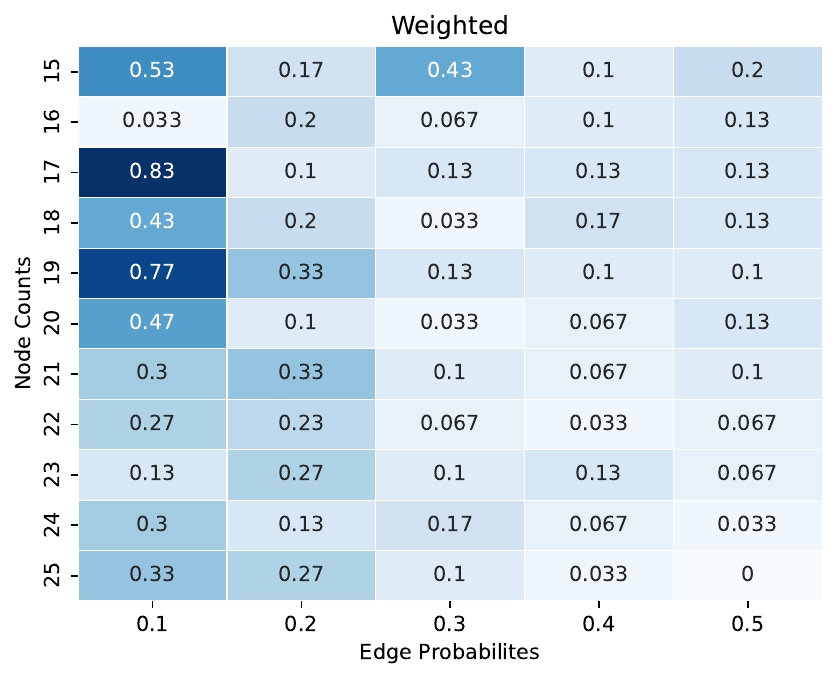}
    \hfill
        \includegraphics[width=0.475\linewidth]{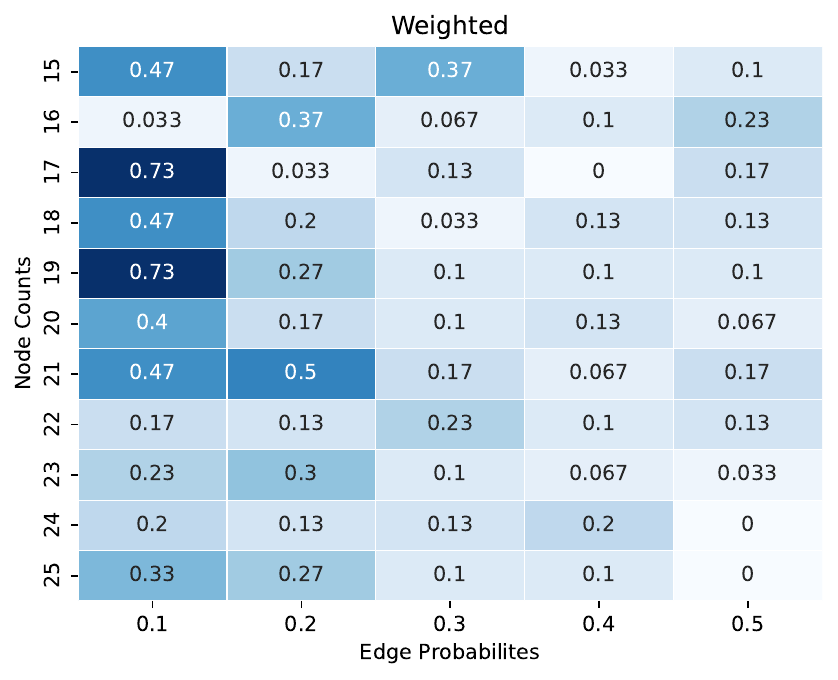}
        \caption{Proportions of cases in which the \gls{qaoa} value is $[95,100)\%$ of \gls{gw} for unweighted (left) and weighted (right) graphs.}   
    \end{subfigure}
    \vskip\baselineskip
    \begin{subfigure}[b]{\textwidth}   
        \centering 
        \includegraphics[width=0.475\linewidth]{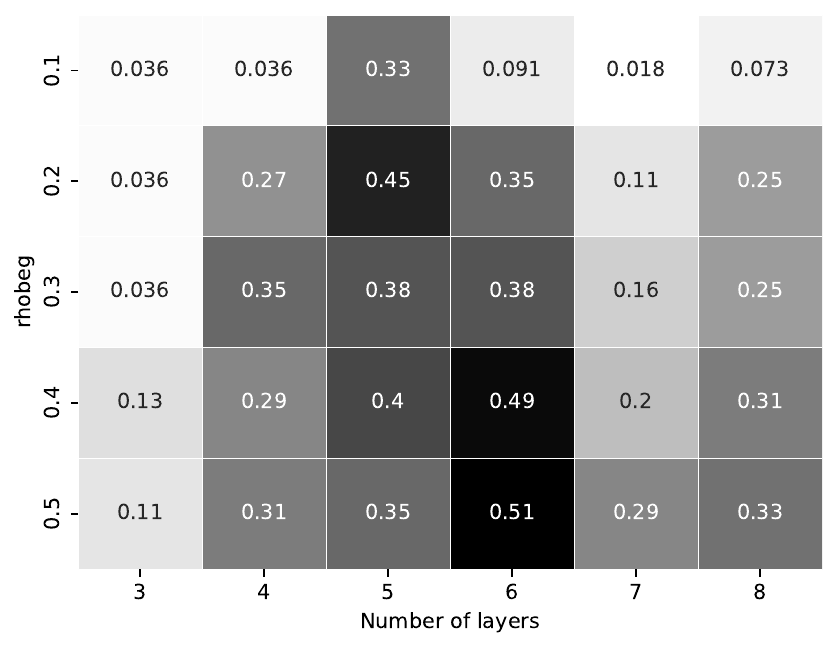}
    \hfill
        \includegraphics[width=0.475\linewidth]{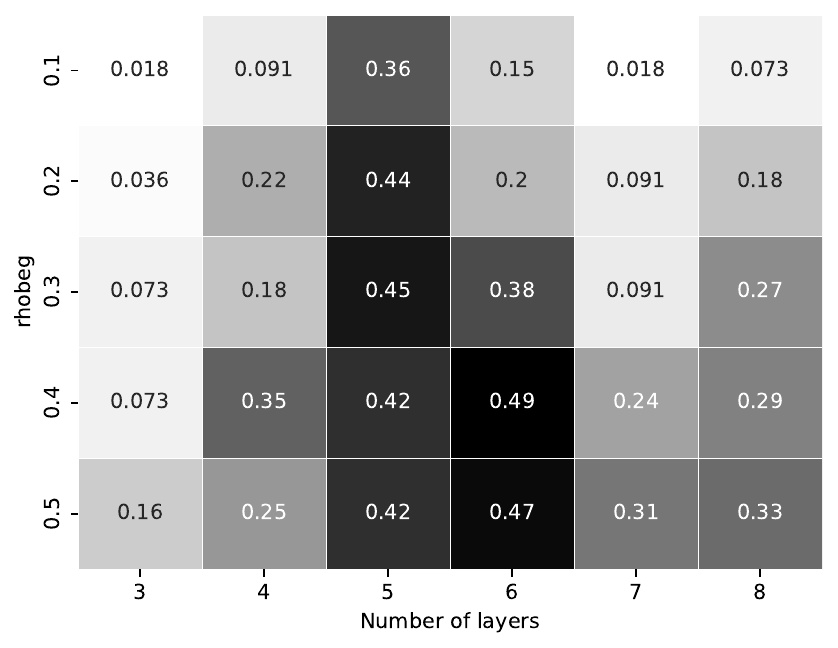}
        \caption{Proportions of cases in the grid search for which \gls{qaoa} is strictly better than \gls{gw} for unweighted (left) and weighted (right) graphs.}   
    \end{subfigure}
    \caption{Results from a grid search over different number of circuit layers and optimization parameters. Weighted and unweighted graphs with different node counts $15$ to $25$ and edge probabilities $0.1$ to $0.5$ have been considered, i.e. 30 grid points.}
    \label{fig:grid_search}
\end{figure*}

\begin{figure}[htbp]
\centerline{\includegraphics[width=0.75\textwidth]{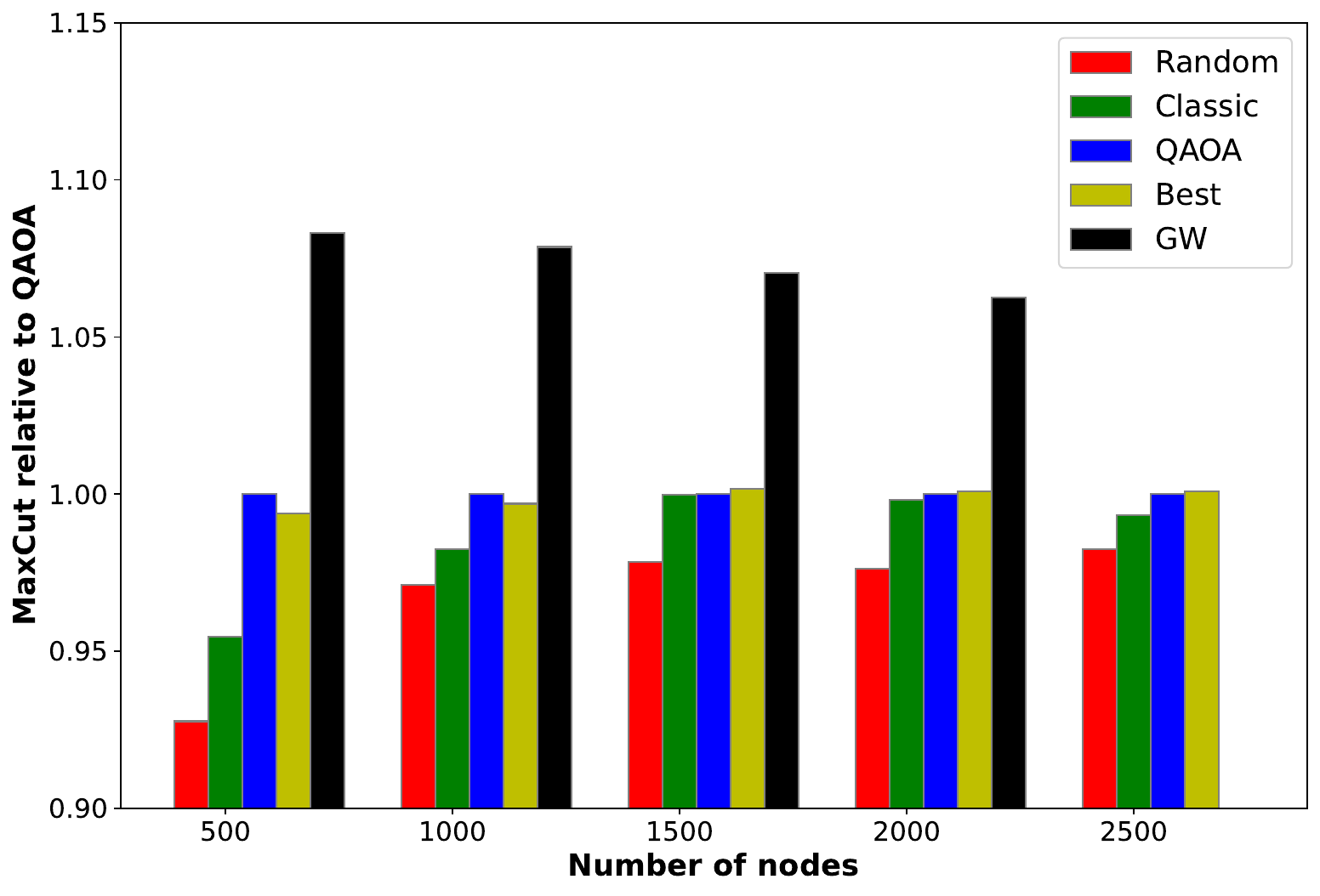}}
\caption{\gls{maxcut} value for unweighted graphs with different node counts and edge probability $0.1$. The sub-graphs from the first partition of the \gls{qaoa2} method are solved either all with \gls{qaoa} (blue), or all classically with \gls{gw} (green), or with the best among the two (yellow). The result from \gls{gw} applied to the original graph is reported as well (black) together with a random partition solution (red) from the \texttt{NetworkX} algorithms suite. Data is relative to the \gls{qaoa} solution.}
\label{fig:comparison}
\end{figure}

To identify an advantage of \gls{qaoa} over \gls{gw} for the solution of the sub-graphs in \gls{qaoa2}, a grid search over a number of circuit layers $p\in\{3,4,\ldots,8\}$ from Eq.~\ref{eq:QAOA_Ansatz} and values of the initial change to the variables for the COBYLA optimizer $\rm rhobeg \in \{0.1, 0.2, 0.3, 0.4, 0.5\}$ has been considered, where the employed Erdős–Rényi graphs are generated with the \texttt{networkx} Python package. The number of iterations is chosen to be linearly dependent on $p$ and ranges from $30$ to $100$ steps. The grid search is applied to every graph in a set with
node counts ranging from $15$ to $25$ and edge probabilities from $0.1$ to $0.5$. A graph instance with uniform edges and one with edge weights randomly chosen in $[0,1]$ is created for every node count and edge probability. The proportions of cases in 
which \gls{qaoa} yields strictly larger MaxCut values than \gls{gw} are recorded, 
as well as the limiting case where \gls{qaoa} reaches $[95,100)\%$ of the  
\gls{gw} value. When \gls{qaoa} is strictly better than \gls{gw}, a score for the corresponding data point in 
the search grid is increased by $1$ and finally normalized. All results are summarized in Fig.~\ref{fig:grid_search}.
This creates a simple, yet instructive, knowledge base about which type of parameterization of \gls{qaoa} is more suitable for a type of graph or whether a classical solution is better overall.
This knowledge can in turn be used to optimally process a set of sub-graphs resulting from a step in \gls{qaoa2}. The grid search reveals that the \gls{qaoa} has a partial advantage for graphs with small edge connection probabilities and that the most successful parameter combination is $(\rm rhobeg = 0.5, p=6)$. 
For higher $p$-layers, \gls{qaoa} is expected to reach better result using more iterations or better initial parameters. 
In a naive approach, all sub-graphs with small edge probabilities would be simulated with \gls{qaoa} using the most successful parameterization or classically otherwise. 
A more advanced approach could use \gls{ml} to train a predictive model and test the performance 
using the workflow described in Sec.~\ref{subsec:workflow}. The same analysis is repeated for node counts $30$ to $33$ and edge probabilities $0.1$ and $0.2$ and summarized in Table \ref{tab1}.
\begin{table}[htbp]
\caption{Analogous data to Fig.~\ref{fig:grid_search} (a) and (b) for node counts $30$ to $33$ and edge probabilities $0.1$ and $0.2$. The top table shows the proportions of cases in which \gls{qaoa2} is strictly better than \gls{gw} and the bottom table shows the proportions of cases in which the \gls{qaoa2} value is $[95,100)\%$ of \gls{gw}.}
\begin{center}
\begin{tabular}{|c|c|c|c|}
\hline
\textbf{Number of} & \textbf{Weighting} & \multicolumn{2}{|c|}{\textbf{Edge Probability}} \\
\cline{3-4} 
\textbf{nodes}                         &                    &\textbf{0.1}&\textbf{0.2}\\
\hline
\hline
30                       &    yes             &      0.1         &       0.1        \\
\cline{2-4}
                        &    no             &        0.167       &         0      \\
\hline
31                      &    yes             &    0.267           &        0.033       \\
\cline{2-4}
                        &    no             &       0        &        0.067       \\
\hline
32                       &    yes             &       0.1        &       0.033        \\
\cline{2-4}
                        &    no             &        0.1       &          0     \\
\hline
33                       &    yes             &      0.033         &        0.033       \\
\cline{2-4}
                        &    no             &       0.167        &          0.033     \\
\hline
\hline
30                       &    yes             &       0.133       &       0.2        \\
\cline{2-4}
                        &    no             &        0.33       &       0.1        \\
\hline
31                      &    yes             &       0.1        &       0.1        \\
\cline{2-4}
                        &    no             &       0.2        &        0.033       \\
\hline
32                       &    yes             &      0.167         &        0.067       \\
\cline{2-4}
                        &    no             &        0.167       &         0.133      \\
\hline
33                       &    yes             &       0.067        &        0.167       \\
\cline{2-4}
                        &    no             &       0.2        &        0.067       \\
\hline
\end{tabular}
\label{tab1}
\end{center}
\end{table}
The occurrences of \gls{qaoa} being strictly better than \gls{gw} are less frequent and no 
distinct point in the parameter grid can be identified as for the lower node counts. Though, a high 
\texttt{rhobeg} or a high number of layers seem more successful. The simulation of \gls{qaoa}
for $33$ qubits takes approximately 10 minutes on 512 compute nodes for $p=8$.
The \gls{qaoa2} is finally applied to a set of larger unweighted graphs with node counts $\{500, 1000, 1500, 2000, 2500\}$ and edge probability $0.1$. The sub-graphs from the first partition are analyzed with the same parameter grid search from before, and the \gls{qaoa} solution with the highest \gls{maxcut} value is stored along with a \gls{gw} solution for that sub-graph, where the best result among the two is recorded. In case of further iterations in the \gls{qaoa2} method, the classical solution is chosen. The results are summarized in Fig.~\ref{fig:comparison} together with the \gls{gw} solution of the original graph and a random partition solution with \texttt{approximation.maxcut} from the \texttt{NetworkX} algorithms suite. 
The \gls{gw} method applied to the full graph is superior compared to the other schemes up to $2000$ nodes, where abnormal terminations are encountered, and diminishes steadily compared to \gls{qaoa2} for larger node counts. The abnormal termination of \gls{gw} beyond $2000$ nodes seems to be related
to the triplet representation in the \texttt{Eigen} library used by \texttt{cvxpy} and not yet to memory consumption. Choosing the best solution among \gls{qaoa} and \gls{gw} for the sub-graphs 
yields slightly better results compared to solving all sub-graphs either with \gls{qaoa} or \gls{gw} in \gls{qaoa2}. Finally, all methods are better than a random cut. The overhead incurred by the coordination of the various sub-graph solutions is minimal and overall an almost ideal scaling is
achieved. 
%
%
\newpage
\section{Conclusion}
The \gls{qaoa} method has been compared to the \gls{gw} algorithm for the \gls{maxcut} problem of graphs with node counts up to $33$ and equivalent number of simulated qubits. For certain parameters, the \gls{qaoa} can outperform its classical counterpart. This knowledge base motivates 
the selection of a most appropriate method for the solution of sub-graphs in the \gls{qaoa2} method.
Experimental results show that the best choice for a sub-graph does not yield significantly better
results compared to a purely classical or (simulated) quantum mechanical solution and is still substantially worse than the \gls{gw} method for the entire graph. However, the \gls{gw} advantage diminishes for larger node counts, which leads to the assumption that \gls{qaoa2} could have an effective advantage at some point. This motivates the investigation of other graph types and partitions including more statistics. Also, considering a larger number of amplitudes in the resulting state vectors is expected to significantly improve the \gls{qaoa} results. The presented simulation infrastructure based on SLURM, which constitutes another research interest, can be further exploited for this analysis. In a future work, the extension to heterogeneous SLURM jobs as well as the use of \gls{ml}
methods for the selection of the best method will be investigated.

\section*{Acknowledgment}
We want to thank Nadav Yoran and Adam Goldfeld from Classiq Technologies as well as Alfio Lazzaro, Jess Jones, and David Brayford from HPE for fruitful discussions.
%
%
%
\bibliographystyle{IEEEtran}
\bibliography{IEEEabrv, paper}
\end{document}